# An RPC-PET brain scanner demonstrator: first results

Paulo Fonte[1,2,3,*], Luís Lopes[1], Rui Alves[1], Nuno Carolino[1], Paulo Crespo[1,4], Miguel Couceiro[1,3], Orlando Cunha[1], Nuno Dias[1], Nuno C. Ferreira[2], Susete Fetal[1,3], Ana L. Lopes[1,4], Jan Michel[5], Jorge Moreira[1], Américo Pereira[1], João Saraiva[1], Carlos Silva[1], Magda Silva[2], Michael Traxler[6], Antero Abrunhosa[2,7], Alberto Blanco[1], Miguel Castelo-Branco[2,7], Mário Pimenta[1,8]

[1] LIP - Laboratory of Instrumentation and Experimental Particle Physics, Coimbra, Portugal
[2] ICNAS Pharma, Institute for Nuclear Sciences Applied to Health, University of Coimbra - Coimbra, Portugal
[3] Coimbra Polytechnic – ISEC, Coimbra, Portugal
[4] Department of Physics, University of Coimbra, 3004-516 Coimbra, Portugal
[5] Goethe Univ, Inst. Kernphys, D-60438 Frankfurt, Germany
[6] GSI Helmholtzzentrum für Schwerionenforschung GmbH, D-64291 Darmstadt, Germany
[7] Coimbra Institute for Biomedical Imaging and Translational Research (CIBIT), University of Coimbra - Coimbra, Portugal
[8] Instituto Superior Técnico (IST), Universidade de Lisboa (UL), Portugal

* Corresponding author: Instituto Superior de Engenharia de Coimbra, Rua Pedro Nunes, 3030-199, Coimbra, Portugal.

## Abstract

We present first results from a Positron Emission Tomography (PET) scanner demonstrator based on Resistive Plate Chambers and specialized for brain imaging. The device features a 30 cm wide cubic field-of-view and each detector comprises 40 gas gaps with 3D location of the interaction point of the photon. Besides other imagery, we show that the reconstructed image resolution, as evaluated by a hot-rod phantom, is sub-millimetric, which is beyond the state-of-the-art of the standard PET technology for this application.

# 1 Introduction

Positron Emission Tomography (PET) is a functional medical imaging technique with important applications in the diagnosis and investigation of some of the most important diseases of the central nervous system such as neurological (Epilepsy, Alzheimer's, Parkinson's, Huntington), psychiatric disorders (depression, schizophrenia), stroke (cerebrovascular accident) and neuro-oncology.

The use of this technique has, however, been limited by poor spatial resolution (typically a few millimeters) when compared to other brain imaging techniques such as Magnetic Resonance Imaging (MRI) or Computed Tomography (CT), which, lacking the sensitivity and the specificity of PET, present better spatial resolution (less than 1 mm). Specialized PET brain scanners are a very active area of research (e.g. [1]).

The converter-plate approach to gamma-ray detection in PET [2] fits particularly well the naturally layered structure of multigap Resistive Plate Chambers (RPC) [3][4]. While certainly unable to reach the detection efficiency typical of scintillator-based systems, this approach (RPC-PET), allowing the precise 3D determination of the interaction point of the photons, promises spatial resolutions below 0.5 mm FWHM [5] with applications in animal and human precision PET imaging. The possibility of economic coverage of large areas along with a time resolution on the level required for time-of-flight PET (TOF-PET) [6] may prove advantageous for total-body human PET [7][8].

In this paper we report first results from an RPC-PET scanner demonstrator dedicated to the human brain and aimed at the best possible position resolution. The sensitivity was limited by the material resources available and it was not a major design goal.

This class of equipment has the potential to change the paradigm in the diagnosis and investigation of diseases of the central nervous system by allowing, for instance, to resolve small brain structures such as the striatum, amygdala and thalamic subnuclei involved in neuropsychiatric diseases. On the other hand, the high spatial resolution of the system may play an important role in the characterization of vascular injuries, improving diagnosis and guiding therapeutics, and in the detection and staging of central nervous system tumors, allowing a better planning of surgery and radiotherapy in cancer patients.

# 2 Description of the scanner

The scanner was constituted by four identical detecting heads each with an active area of $30\times30$ cm$^2$, disposed to form a 30 cm wide cubic field-of-view, as shown in Figure 1 a). The relative solid angle subtended by the detectors as viewed from the central point was, therefore, close to $\Omega_r = 66\%$.

Each head was equipped with 8 multigap RPCs of the type already described in [5], each comprising 5 gas gaps of 0.35 mm and 6 glass plates with a thickness of 0.28 mm. In this prototype the heads may eventually be upgraded to 16 RPCs, more than tripling the sensitivity to the photon pairs.

The heads were flushed with a mixture of gaseous R134a + 2% SF$_6$.

Each RPC was readout on both sides by thin printed circuit board electrodes equipped with 48 readout strips of 6.4 mm pitch, each side reading one coordinate (X or Y). The stacking (see Figure 1 b)) was repeated so that each electrode sensed two adjacent

RPCs, comprising in total 4 X-electrodes and 5 Y-electrodes. Each X-electrode also included 10 outputs for the timing/trigger amplifiers (see [5]).

Simulations [9] suggest that the material budget of the readout electrodes has only a minor negative effect on the detection efficiency of the heads.

The front-end electronics, comprising timing/trigger amplifiers and charge amplifiers, is housed on the shielded back side of the detecting heads, as it can be seen Figure 1 a).

The 10-channel timing/trigger amplifier boards were custom-built and based on two stages of amplification by the SPF5043Z MMIC followed by a MAX9601 comparator. The gain-bandwidth product was close to 60 GHz and the peak noise level at the comparators was close to 50 mV, corresponding to 20 µA at the input. The threshold of each channel could be individually adjusted and the boards were equipped with a wired-OR output for trigger purposes.

The readout strips were connected to 24-channel custom bipolar charge amplifier boards with integration time constant of 20 µs and gain of 250 mV/pC. The pulses were digitally processed after digitization.

The front-end electronics was powered independently in each head by a linear AC/DC converter.

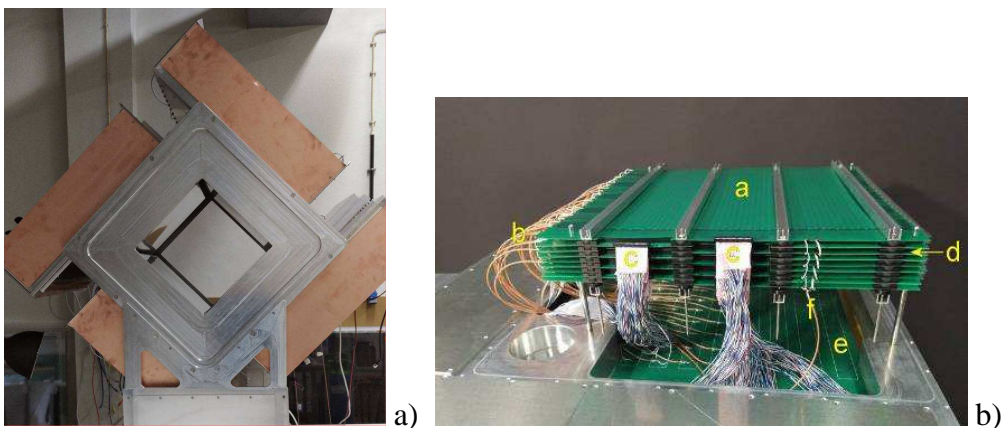

Figure 1 – a) View of the arrangement of the detecting heads, forming a 30 cm wide cubic field-of-view. b) Detector stack, showing the readout electrodes (a), the connections for the timing/trigger amplifiers (b), the connections to the charge amplifiers (c), the RPCs (d) and a spare space for future upgrade (e).

The data acquisition system was based on the TRB family of electronics [10] and comprised 16 48-channel streaming ADCs with 40 MHz sampling rate and in-board digital pulse processing ("ADC AddOn"). It also included 64 channels of TDC [11] with 10 ps bin size and a flexible central trigger system. The data was sent from each ADC via a 1GbE connection to a central commercial switch that consolidated the data streams into two 10GbE connections fed to a central event-building server.

Each head was supplied by two bipolar high-voltage units mounted on the head, each supplying up to ±9 kV under the control of the I2C-based slow control system. This system also controlled the threshold settings, the gas system and the monitoring sensors.

# 3 Results

## 3.1 Charge distribution

The charge distribution observed from coincidences between 511 keV photons is shown in Figure 2. It can be seen that when a voltage of 16.8 kV is applied to the RPC, corresponding to an electric field of 96 kV/cm in each gas gap, it appears a shallow peak in the distribution, detached from the cutoff caused by the threshold of the trigger amplifiers.

To estimate the fraction of events lost in these conditions, in the right hand side panel we fitted the cumulative distribution around the position of the distribution peak with a cubic polynomial and extrapolated the polynomial to zero charge, which yields a value close to 1.2. Therefore it is estimated that the cutoff reduces the efficiency to $1/1.2 = 83\%$ of its intrinsic value.

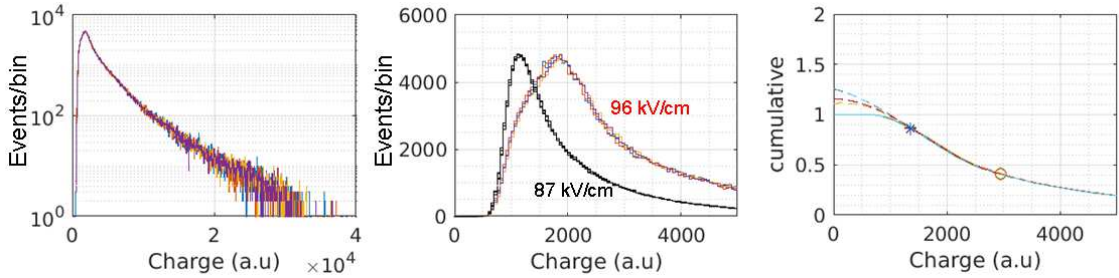

Figure 2 – Left panel: four charge distributions measured at 96 kV/mm in the X and Y electrodes of two RPCs on different heads. Central panel: detail of the same distributions close to the origin, along with, for comparison, a similar distribution for a lower voltage (black curves), evidencing the cut from the trigger amplifiers. Right panel: cumulative charge distribution and a cubic polynomial fit (dashed lines) to the peak region (between the symbols), extrapolated to zero charge.

## 3.2 Sensitivity and detection efficiency

The central point sensitivity[*] (CPS) of the system measured per the NEMA-NU4-2008 standard as a function of the applied voltage is shown in Figure 3. The highest voltage applied (2×8.4 kV) corresponds to the charge spectrum presented in Figure 2, yielding a value of 0.092%. It is apparent that there is little sensitivity to be gained beyond this voltage.

In principle, the detection efficiency, $\varepsilon$, the relative solid angle and the CPS are related by $CPS = \Omega_r \varepsilon^2$, from which one can deduce an experimental detection efficiency $\varepsilon = 3.7\%$ ($\Omega_r = 66\%$). This value is supposed to be affected by the cut in the charge spectrum estimated in the previous section to amount to 83%, resulting in an intrinsic detection efficiency of 4.5%.

It should be noted that in this demonstrator the stack comprised only 8 RPCs per head. In Figure 3, left panel, it is presented the extrapolation of the observed CPS to a larger number of RPCs by normalizing the simulated detection efficiency [9] to the observed CPS and keeping the same trend as given by simulation. It suggests that by tripling the amount of RPCs per head, which we consider feasible, a 6–fold increase in CPS may be possible.

---

[*] The probability of detection of a photon pair emitted from the centre of the tomograph.

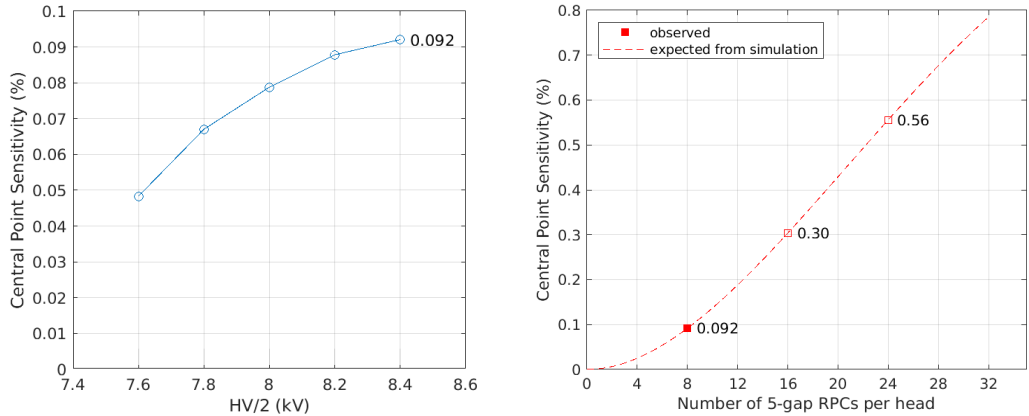

Figure 3 – Left panel: Central point sensitivity as a function of the applied voltage, measured in accordance with the NEMA NU4-2008 standard. Right panel: the observed CPS is extrapolated to a larger number of RPCs by following the trend obtained by simulation.

## 3.3 Image resolution

The image resolution was evaluated by imaging a hot-rod or "Derenzo" phantom, filled with a liquid radiotracer containing $^{18}$F. The phantom had 6 sectors with several inter-disc distances to establish the resolution limit. The initial activity in the phantom was 500 µCi and 5 million events were recorded.

The data was reconstructed by an iterative Maximum Likelihood – Expectation Maximization (MLEM) algorithm. The result can be seen in Figure 4, showing the 1.0 mm slice well resolved and the 0.95 mm slice partially resolved, therefore demonstrating a sub-millimetric image resolution.

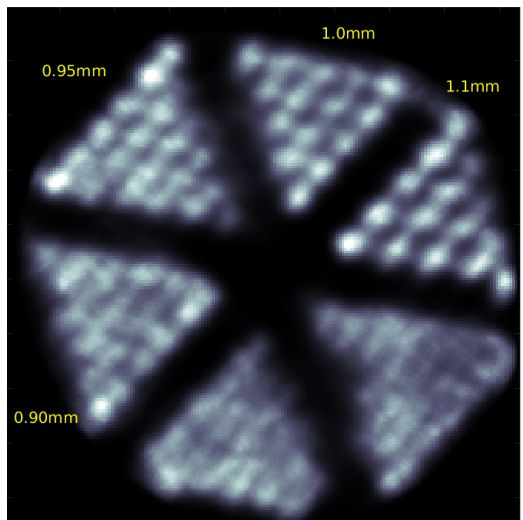

Figure 4 –Image of a hot-rod phantom, showing the 1.0 mm slice well resolved and the 0.95 mm slice partially resolved, demonstrating a sub-millimetric image resolution.

## 3.4 Brain phantom

To address the capability of imaging realistic objects and activity levels we took data with a brain and striatum phantom [12] that comprised a fillable brain cavity with a volume of 1260 mL and separately fillable left and right caudate and putamen nuclei with 5.4 mL and 6.0 mL, respectively.

All cavities were filled with a liquid radiotracer containing $^{18}$F. Data was taken over a 4 h period (~2 decay half-lives) with average activity concentrations of 7.9 kBq/mL on the brain cavity and about 8-fold larger on the striatum cavities (63 kBq/mL), having been collected 37 million events.

Although the collecting time was long, it should be noted that the second half-live only contributed with 1/3 of the events and that on a more definitive scanner (this is only a demonstrator) the sensitivity could be strongly improved, as discussed in section 3.2.

In Figure 5 are shown images of the full brain, the striatum and a detail view evidencing the separation between the caudate and putamen cavities. In the first image it is also visible a small $^{22}$Na source that was placed externally on the forehead of the phantom for quality assurance purposes. The image reconstruction algorithm used was the same as for the resolution phantom.

It is apparent that noise free images are possible, with good resolution, evidenced by the visible separation between the caudate and putamen cavities, which are externally touching.

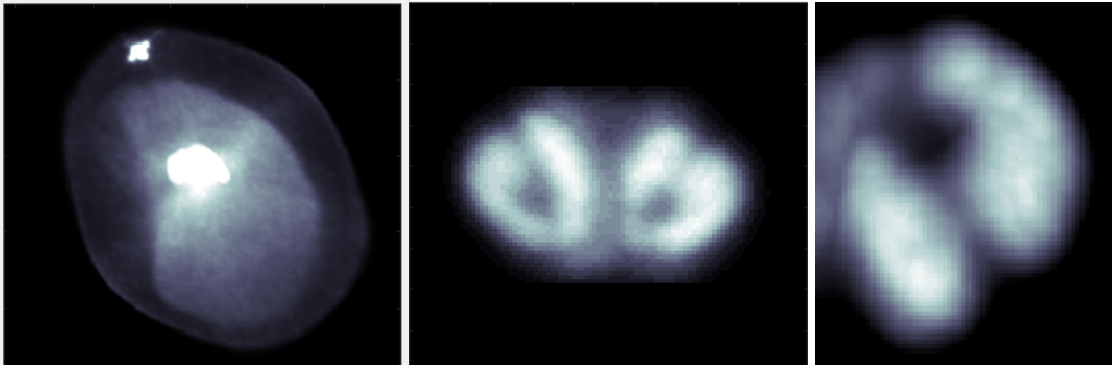

Figure 5 –Images of a brain and striatum phantom [12]. From left to right: full brain, including a small $^{22}$Na source that was placed externally on the forehead of the phantom; the striatum nuclei; a detail view evidencing the separation between the caudate and putamen cavities.

## 4 Conclusion

We demonstrated the viability of the application of the RPC-PET approach to specialized human brain PET scanning.

First results include the demonstration of a sub-millimetric image resolution, which is beyond the state-of-the-art, and the obtention of detailed images of the striatum nuclei in a brain phantom.

## 5 References

[1] L. Moliner, M.J. Rodriguez-Alvarez, J.V. Catret, A. Gonzalez, V. Ilisie, J.M. Benlloch, NEMA Performance Evaluation of CareMiBrain dedicated brain PET

and Comparison with the whole-body and dedicated brain PET systems, Scientific Reports, 9 (2019).

[2] J.E. Bateman, X-RAY AND GAMMA-IMAGING WITH MULTIWIRE PROPORTIONAL-COUNTERS, Nuclear Instruments & Methods in Physics Research Section A 221 (1984) 131-141.

[3] E.C. Zeballos, I. Crotty, D. Hatzifotiadou, J.L. Valverde, S. Neupane, M.C.S. Williams, A. Zichichi, A new type of resistive plate chamber: The multigap RPC, Nuclear Instruments & Methods in Physics Research A, 374 (1996) 132-135.

[4] A. Blanco, V. Chepel, R. Ferreira-Marques, P. Fonte, M.I. Lopes, V. Peskov, A. Policarpo, Perspectives for positron emission tomography with RPCs, Nuclear Instruments & Methods in Physics Research Section A, 508 (2003) 88-93.

[5] P. Martins, A. Blanco, P. Crespo, M.F.F. Marques, R.F. Marques, P.M. Gordo, M. Kajetanowicz, G. Korcyl, L. Lopes, J. Michel, M. Palka, M. Traxler, P. Fonte, Towards very high resolution RPC-PET for small animals, Journal of Instrumentation, 9 (2014).

[6] L. Lopes, R.F. Marques, P. Fonte, L. Hennetier, A. Pereira, A.M.S. Correia, Ceramic high-rate timing RPCs, Nuclear Physics B-Proceedings Supplements, 158 (2006) 66-70.

[7] V. Nadig, K. Herrmann, F.M. Mottaghy, V. Schulz, Hybrid total-body pet scanners-current status and future perspectives, European Journal of Nuclear Medicine and Molecular Imaging, 49 (2022) 445-459.

[8] M. Couceiro, P. Crespo, R.F. Marques, P. Fonte, Scatter Fraction, Count Rates, and Noise Equivalent Count Rate of a Single-Bed Position RPC TOF-PET System Assessed by Simulations Following the NEMA NU2-2001 Standards, IEEE Transactions on Nuclear Science, 61 (2014) 1153-1163.

[9] Ana Luísa Lopes, Miguel Couceiro, Paulo Crespo and Paulo Fonte, Optimization through Monte Carlo Simulations of a novel High-Resolution Brain-PET System based on Resistive Plate Chambers, IEEE Nuclear Science Symposium / Medical Imaging Conference Record (NSS/MIC), Piscataway, NJ, USA, 2021 (doi: 10.1109/NSS/MIC44867.2021.9875731)

[10] "TRB Colaboration", trb.gsi.de

[11] C. Ugur, S. Linev, J. Michel, T. Schweitzer, M. Traxler, A novel approach for pulse width measurements with a high precision (8 ps RMS) TDC in an FPGA, Journal of Instrumentation, 11 (2016)

[12] Striatal Phantom manufactured by Radiology Support Devices, 1904 East Dominguez Street, Long Beach, CA 90810

# 6 Acknowledgement

This work was financed by the European Union via the programs PT2020 and COMPETE2020 (project POCI-01-0247-FEDER-039808).